# Efficient Genomic Interval Queries Using Augmented Range Trees

**Chengsheng Mao[1], Alal Eran[2, 3], Yuan Luo[1, *]**

[1]Department of Preventive Medicine, Northwestern University Feinberg School of Medicine, Chicago, IL, USA;

[2]Department of Biomedical Informatics, Harvard Medical School, Boston, MA, USA;

[3]Department of Life Sciences, Ben Gurion University of the Negev, Israel;

*Corresponding author: yuan.luo@northwestern.edu

## ABSTRACT

Efficient large-scale annotation of genomic intervals is essential for personal genome interpretation in the realm of precision medicine. There are 13 possible relations between two intervals according to Allen's interval algebra. Conventional interval trees are routinely used to identify the genomic intervals satisfying a coarse relation with a query interval, but cannot support efficient query for more refined relations such as all Allen's relations. We design and implement a novel approach to address this unmet need. Through rewriting Allen's interval relations, we transform an interval query to a range query, then adapt and utilize the range trees for querying. We implement two types of range trees: a basic 2-dimensional range tree (2D-RT) and an augmented range tree with fractional cascading (RTFC) and compare them with the conventional interval tree (IT). Theoretical analysis shows that RTFC can achieve the best time complexity for interval queries regarding all Allen's relations among the three trees. We also perform comparative experiments on the efficiency of RTFC, 2D-RT and IT in querying noncoding element annotations in a large collection of personal genomes. Our experimental results show that 2D-RT is more efficient than IT for interval queries regarding most of Allen's relations, RTFC is even more efficient than 2D-RT. The results demonstrate that RTFC is an efficient data structure for querying large-scale datasets regarding Allen's relations between genomic intervals, such as those required by interpreting genome-wide variation in large populations.

# 1 INTRODUCTION

Annotating functional elements in genomic datasets is fundamental for understanding genome biology, interpreting genomic variation, and advancing precision medicine. Genomic features, such as genes, exons, or regulatory regions, can be represented as genomic intervals, comprised of a chromosome ID with a start and an end position. Genomic intervals serve to anchor numerous diverse genomic datasets and their experimental results on a common basis, thereby facilitating their comparison and integration. With the advance of next-generation sequencing, multiple online resources including the UCSC genome browser [1] and the Encyclopedia of DNA Elements (ENCODE) project [2] provide billions of interval-based genomic annotations. Sifting through a large number of genomic intervals often involves identifying the set of intervals satisfying certain interval relations with query genomic intervals. This is a challenging task due to the extremely large number of genomic intervals present and due to the multiple different relations that can hold between genomic intervals. Most existing studies focus on the overlapping relations between genomic intervals [1, 3-9]. However, more refined relations between genomic intervals can also be suggestive of relations between corresponding genomic annotations. For example, a putative promoter that regulates transcription of a particular gene is located in the vicinity of the transcription start sites of that gene, and more specifically on the same strand, and upstream of the gene. Developing efficient methods for identifying the set of genomic intervals satisfying the often complex relations with a genomic interval of interest is a major unmet need that is crucial for biomedical discoveries.

In many cases, we need to know all the intervals in a large set that satisfy a certain relation with a given interval. This is known as an interval query problem. Given a query interval $q$, a set of data intervals $S$, and one relation $r$, an interval query is to retrieve all the intervals $x \in S$, such that the relation $x\ r\ q$ holds. In interval query, the most common and widely studied problem is the intersection query problem where the relation $r$ refers to the intersection relation. Usually, two intervals $a$ and $b$ intersect if and only if $a.start \leq b.end$ and $a.end \geq b.start$ where for example, $a.start$ denotes the start position of interval $a$. However, in many cases, the intersection relation can be rather coarse, and the relative position of the overlapping genomic intervals may also be of interest. Figure 1 illustrates such an example of interval queries regarding four different types of intersections including: overlapping from the front (*o*), overlapping from behind (*oi*), contains (*di*), and contained in (*d*).

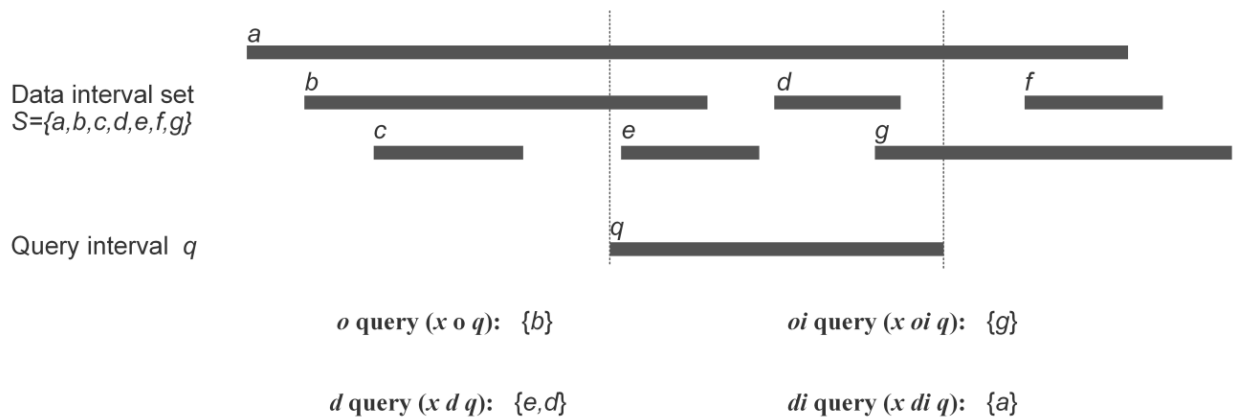

Figure 1  An example of interval queries regarding four different intersection relations, i.e., overlapping from the front (*o*), overlapping from behind (*oi*), contains (*di*) and contained in (*d*). The four types of interval queries are based on the query interval q and the data interval set S.

Figure 1 is a simple interval query example with a small data interval set where the result set can be obtained by comparing each interval in the set with the query interval one by one. However, in practical applications, the brute-force enumeration scales poorly for large datasets in terms of efficiency. Developing efficient methods for interval queries regarding more refined relations is also crucial, e.g., the putative promoter example.

Previous studies on genomic interval query made use of interval trees [10], binning approaches (based on R-trees) [1, 6, 7], nested containment lists [4, 8] or linear sweeps [3, 9], but mostly focused on intersection queries. Seok *et al.* [5] reviewed a number of interval query algorithms from the above categories and analyzed their time complexities on intersection queries: most of them cannot achieve $O(log\ n\ +\ k)$ time, where *n* is the size of interval set and *k* is the number of result intervals. Though interval tree algorithms can achieve $O(log\ n\ +\ k)$ time on intersection queries, however, theoretical analysis showed that conventional interval tree based algorithms have sub-optimal speed (time complexity larger than $O(log\ n\ +\ k)$ in general) for more refined interval relations in Allen's algebra [11]. In this paper, we propose query rewriting and adapt the range trees to design and implement an efficient interval query method that can achieve the optimal $O(log\ n\ +\ k)$ time complexity for interval queries regarding all Allen's interval relations.

## 2   OBJECTIVES

Efficient queries for genomic intervals that have a certain relation with a given interval are essential for various bioinformatic applications, especially for large genomic datasets. According to Allen's interval algebra, there are 13 possible relations between two intervals, 11 out of which are associated with the intersection, the other two are associated with non-intersection. An interval query regarding one of the two non-intersection relations can be recast as a certain interval query regarding one of Allen's intersection relations, so we only consider the interval queries regarding Allen's intersection relations. Though the coarse intersection query is widely studied, and the studies have made some achievements, interval queries regarding more refined relations are also of interest. Unfortunately, existing interval query methods designed for the coarse intersection relation cannot be efficiently extended to interval queries regarding more refined relations in Allen's algebra. Our objective is to improve the interval query efficiency regarding refined relations in Allen's algebra using query rewriting and the range tree data structure.

## 3    MATERIALS AND METHODS

We applied Allen's interval algebra to refine the relation between two genomic intervals into 13 categories. To efficiently retrieve all the genomic intervals satisfying a certain Allen's relation with a given genomic interval from a large dataset, we regarded an interval as a 2-dimensional point and transformed the interval query problem to the range query problem by rewriting the definition of Allen's interval relations. We then applied the range tree data structure and the corresponding query algorithm to perform efficient range queries. Besides the conventional 2-dimensional range tree (2D-RT), we also augmented the range tree structure to improve the query efficiency using the technique of fractional cascading and name the implementation as range tree with fractional cascading (RTFC). The current state-of-the-art interval tree (IT) algorithm was also implemented as a baseline. We tested interval query efficiency of the above algorithms with ENCODE [2] genomic annotation intervals as the data interval dataset, and Genome Aggregation Database (gnomAD) [12] variant intervals as the query set. We have made our code publicly available at GitHub upon the acceptance of the paper.

## 3.1 Allen's Interval Algebra

In 1-dimensional cases, an interval is usually defined by two numbers corresponding to the start and the end, where the end is supposed to be greater than the start. In this paper, we use $[x, y]$ to denote an interval with start $x$ and end $y$. Based on the three relations between two numbers, i.e., greater, equal and less, Allen [13] proposed 13 relations between two temporal intervals that are distinctive, exhaustive and qualitative. Distinctive and exhaustive because each pair of definite intervals must be described by one and only one of the relations; qualitative because no numeric spans are considered. The relations between intervals and the operations based on them form Allen's interval algebra. Though Allen's interval algebra was originally proposed for temporal intervals, it applies to spatial intervals such as genomic intervals. If we consider two intervals, $a = [x, y]$ and $q = [x', y']$, the 13 relations between them can be defined and illustrated in Table 1. From Table 1, Allen's interval algebra provides a more refined interval relation category method, based on which the intersection relation consists of 11 Allen's interval relations (i.e., *o, oi, d, di, s, si, f, fi, m, mi* and =).

For an interval query regarding relation '<' or '>', there usually are a very large number of result intervals. In practice, only a subset of these result intervals are of interest, and the query can be regarded as a *d* query. For example, to find a putative promoter in the upstream of the gene with interval $[x', y']$ requires a '<' query. Since a promoter that regulates transcription of the gene is located in the vicinity (e.g., l bases) of the transcription start sites of that gene (i.e., $x'$), we only need to perform the *d* query regarding $[x' - l, x']$. Analogously, a '>' query also usually takes the form of a *d* query regarding $[y', y' + l]$ in practice. Thus we only consider the 11 intersection related queries.

Table 1 Allen's interval relations and their transformation to the bound range for range query. Note: $a = [x, y]$ is an interval from the data interval set, $q = [x', y']$ is the query interval.

| Symbols | Relation | Illustration | Definition | Rewriting as range query |
|---|---|---|---|---|
| o | a o q | 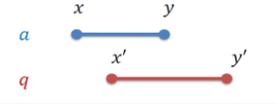 | $x < x' < y < y'$ | $0 < x < x'$ <br> $x' < y < y'$ |
| oi | a oi q | 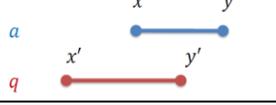 | $x' < x < y' < y$ | $x' < x < y'$ <br> $y' < y < \infty$ |
| d | a d q | 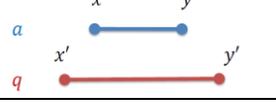 | $x' < x < y < y'$ | $x' < x < y'$ <br> $x' < y < y'$ |
| di | a di q | 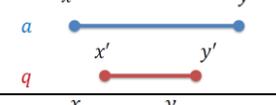 | $x < x' < y' < y$ | $0 < x < x'$ <br> $y' < y < \infty$ |
| m | a m q | 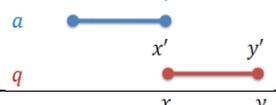 | $x < y = x' < y'$ | $0 < x < x'$ <br> $y = x'$ |
| mi | a mi q | 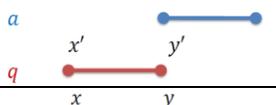 | $x' < y' = x < y$ | $x = y'$ <br> $y' < y < \infty$ |
| s | a s q | 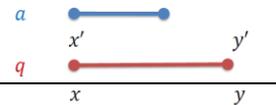 | $x' = x < y < y'$ | $x = x'$ <br> $x' < y < y'$ |
| si | a si q | 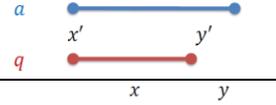 | $x = x' < y' < y$ | $x = x'$ <br> $y' < y < \infty$ |
| f | a f q | 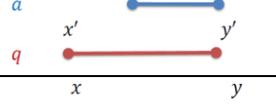 | $x' < x < y = y'$ | $x' < x < y'$ <br> $y = y'$ |
| fi | a fi q | 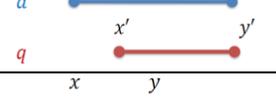 | $x < x' < y' = y$ | $0 < x < x'$ <br> $y = y'$ |
| < | a < q | 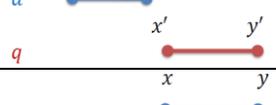 | $x < y < x' < y'$ | $0 < x < x'$ <br> $0 < y < x'$ |
| > | a > q | 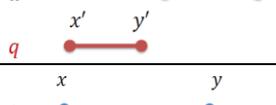 | $x' < y' < x < y$ | $y' < x < \infty$ <br> $y' < y < \infty$ |
| = | a = q | 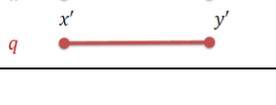 | $x = x' < y' = y$ | $x = x'$ <br> $y = y'$ |

## 3.2 Rewriting Interval query to range query

An interval can be mapped to a 2-dimensional point with the 2 endpoints of the interval as the 2 coordinates of the point, i.e., interval $[x, y]$ corresponds to the point $(x, y)$. Thus, a 2-dimensional range tree can be adapted for interval query by transforming interval relations to relations between 2-dimensional points. Through rewriting the definition of each Allen's interval relation as shown in the last column of Table 1, a satisfying interval can be mapped to a point satisfying a certain range constraint, and thus an interval query problem is transformed to a range query problem, as shown in Figure 2. Figure 2 illustrates the satisfying regions of range queries that are transformed from interval queries with respect to a certain query interval. From Figure 2, it is natural to apply a range tree for interval queries after query rewriting. Note that an interval should have its start less than its end, thus a point associated with an interval must be above the line $y = x$. This is the reason why the lower right area in Figure 2 is invalid for interval query.

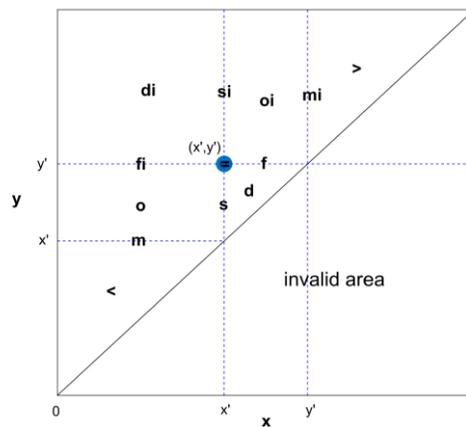

Figure 2 If an interval is associated with a 2-dimensional point, an interval query can be transformed to a range query. The query interval $[x', y']$ is associated with the point $(x', y')$, and the interval query regarding each of Allen's interval relations corresponds to the range query regarding the indicated area (marked in the figure by the corresponding relation symbol, the "=" query exactly corresponds to the point $(x', y')$). The start is less than or equal to the end for an interval, thus, points in the lower right area under the line $y = x$ is invalid to associate an interval with.

## 3.3 Basic range tree

Range trees are originally designed for efficient range queries [14, 15]. In range query problems, range trees are usually applied to query the set of points that lie in a given range, especially in a rectangular area. A *d*-dimensional range tree $RT^d$ on a set of *d*-dimensional points is actually an augmented balanced Binary Search Tree (BST) with the following recursive structure. Each node $v$ uses the first of the *d* coordinates as its key and contains an associated *(d-1)*-dimensional range tree $RT_v^{d-1}$ on the rest *d*-1 coordinates of the points stored in the subtree rooted at $v$. Each node $u$ of $RT_v^{d-1}$ uses the second of the *d* coordinates as its key and contains an associated *(d-2)*-dimensional range tree $RT_u^{d-2}$ on the rest *d-2* coordinates of the points stored in the subtree rooted at $u$. The recursive structure continues analogously as we go through each of the *d* coordinates. Eventually, a 1-dimensional range is exactly a traditional balanced BST on the last coordinate. Generally, the points stored in a range tree are stored in the leaves, and each internal node stores the largest value contained in its left child.

In 1-dimensional cases, a range query is to list all the points that lie in a certain interval, denoted as $[x1, x2]$, from a set of given points. One can use a 1-dimensional range tree to efficiently perform the query by searching for the two endpoints $x1$ and $x2$ respectively and reporting all the points between them. Since the range tree is balanced, the search for $x1$ and $x2$ takes $O(\log n)$ time, where *n* is the number of data points. Reporting all the points between $x1$ and $x2$ needs to traverse all the subtrees between their search paths, and it can be done in linear time. Thus, a range query on the basic 1-dimensional range tree has the time complexity $O(\log n + k)$, where $k$ is the number of result points.

Range queries in *d*-dimensional cases are similar. The main difference is that for *d*-dimensional trees we need to traverse the recursive tree structure as defined above. For example, using the two endpoints of the first coordinate, we identify a set of subtrees $S$ between their search paths (excluding the nodes in the search paths). For the root $v$ of each subtree in $S$, we perform a *(d-1)*-dimensional range query on $RT_v^{d-1}$. To perform a *(d-1)*-dimensional range query on $RT_v^{d-1}$, we identify a set of subtrees $S_v$ using the two endpoints of the second coordinate analogously, then perform a *(d-2)*-dimensional range query on $RT_u^{d-2}$ for the root $u$ of each subtree in $S_v$. We continue with further lower dimensional queries in a recursive manner. Eventually, a series of 1-dimensional range queries will be performed, and the correct points will be reported. Since a *d*-dimensional range query consists of $O(\log n)$ *(d-1)*-dimensional range queries [16, 17], by recursive time complexity analysis, the time required to perform a *d*-dimensional range query is $O(\log^d n + k)$.

## 3.4 Fractional cascading

Fractional cascading (FC) is a technique to speed up the searching for the same value from multiple sequences [18, 19]. With FC, searching for the same value from a number of sets would need only one search for that value from the union of these sets. For example, if we have two arrays of numbers, $A1$ and $A2$ (let $A2 \subseteq A1$), both sorted ascendingly, then FC takes the following steps to search for the minimum values no less than $v$ in $A1$ and $A2$ respectively. This process is also shown in Figure 3.

1. Create an indexing array $I$ from $A1$ to $A2$. The *i*th element in $A1$ (i.e., $A1[i]$) corresponds to the *i*th element in $I$ (i.e., $I[i]$) which is the index of the smallest element in $A2$ no less than $A1[i]$ (-1 if no such elements). We refer to such an index as FC-index.

2. Perform a binary search on $A1$ for $v$ and return the smallest element no less than $v$ in $A1$, say $A1[j]$, and its position $j$.

3. Directly obtain the smallest element no less than $v$ in $A2$, i.e., $A2[I[j]]$.

Through FC, searching for the same value from multiple sequences can be achieved by only one binary search from the union set, and the search results of each subset can be directly obtained; this is more efficient than multiple searches from all the sets.

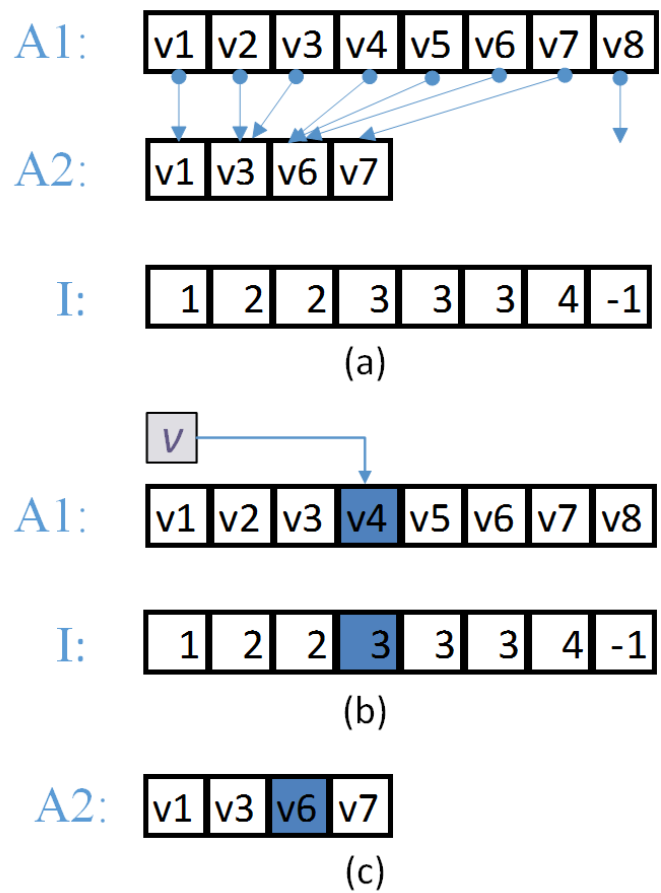

Figure 3 A simple example of search by fractional cascading. Given a value $v$, the query searches for the minimum values no less than $v$ in A1 and A2 respectively. (a) Create the indexing array $I$ from A1 to A2. Refer to such an index as FC-index. (b) Suppose $v \in (v3, v4]$, the binary search for $v$ from A1 will return $v4$ and its position 4, then the corresponding FC-index in A2 is $I[4] = 3$. (c) Directly return $A2[I[4]] = v6$ as the search result from A2.

## 3.5 Range tree with fractional cascading

By the fractional cascading technique, we can improve the query time from $O(log^d n + k)$ to $O(log^{d-1} n + k)$ for range tree [16, 20], and thus a 2-dimensional range query can be done in $O(\log n + k)$ time. Let *x* and *y* indicate the coordinates of the 2 dimensions, we construct a 2-dimensional range tree with fractional cascading (RTFC) using the following steps. The *x*-tree is first constructed on the *x* coordinates of all the points as the basic range tree. Instead of building a *y*-tree $RT_v^1$ in each node $v$, RTFC stores the corresponding points as an array $A(v)$, sorted by *y*-coordinate. In addition, each node $v$ stores two FC-index arrays, containing the FC-indices from $A(v)$ to $A(l(v))$ and $A(r(v))$, where $l(v)$ and $r(v)$ are the left and right children of node $v$ respectively. Since $A(v) = A(l(v)) \cup A(r(v))$, by the fractional cascading technique described in Section 3.4, once we obtain the range query results in $v$, the subsequent query results in its children $l(v)$ and $r(v)$ can directly be obtained through their corresponding FC-index arrays. The query results in its grandchildren are readily obtained, recursively for all descendants until we reach the leaf nodes. Since range queries on RTFC avoid the 1-dimensional range query on *y*-trees, it has the time complexity $O(\log n + k)$ in 2-dimensional cases and $O(log^{d-1} n + k)$ in *d*-dimensional cases.

## 3.6 Data structure implementation

Besides the RTFC, we also implement the traditional 2-dimensional range tree (2D-RT) and the interval tree (IT) as baselines. All the three tree structures are implemented in C++. Unlike IT in which each interval is referenced only once, in RTFC and 2D-RT, each interval is referenced multiple times. To reduce memory consumption, we use a static vector to store the points (i.e., rewritten intervals) and refer to the indices of the point in RTFC and 2D-RT. We design and implement the three tree structures as Figure 4. Our implementation codes are available at https://github.com/yuanluo/rangetree.

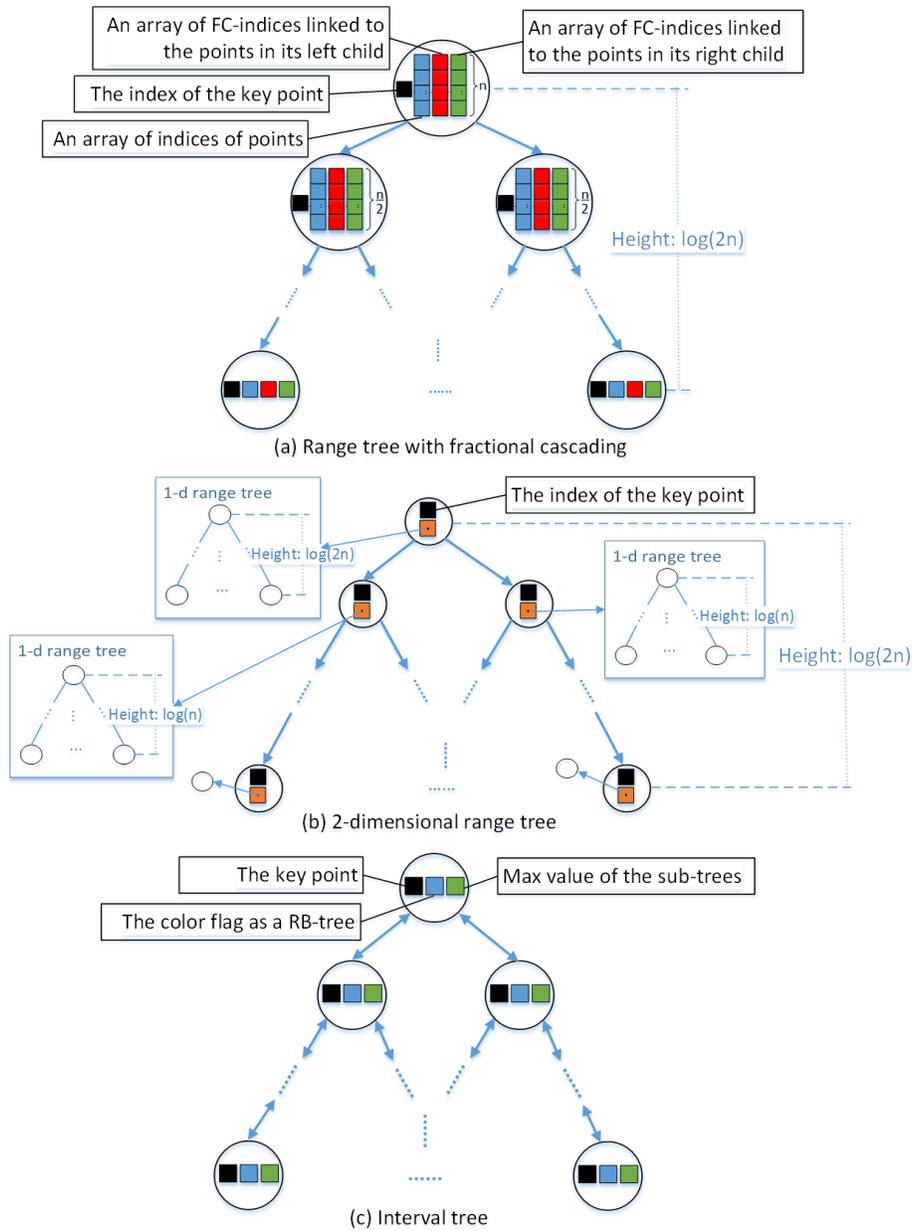

Figure 4 The three tree structures for a data interval set with n intervals. For brevity of illustration, the pointer fields of each node pointing to the left, right or parent node are omitted inside the node, and represented by the corresponding arrows between nodes. (a) The data structure of range tree with fractional cascading (RTFC). (b) The data structure of the basic 2-dimensional range tree (2D-RT). (c) The data structure of interval tree (IT) implemented as an augmented RB-tree (red-black tree) with n nodes.

## 3.7 Dataset

As we move closer to practicing precision medicine, one of the main challenges remains the interpretation of noncoding genomic variants [21, 22]. Since the vast majority of the human genome is noncoding, the vast majority of human variation is noncoding. To assess the functionality of noncoding regions and enable personal noncoding variant interpretation, the ENCODE project has systematically identified functional genomic intervals in the human genome at scale [2]. These include transcription factor binding sites, chromatin structures, and histone modification sites. Here we demonstrate the ability of RTFC to rapidly annotate noncoding variants in these ENCODE regions, thereby facilitating timely personal genome interpretation.

For each of the 10,791 bed files representing high-quality ENCODE ChIP-seq data (supplementary material), we extract the "chromStart" and "chromEnd" fields (start and end positions of a region in a chromosome) to construct intervals. We then categorize all the intervals from all the bed files by the chromosome (i.e., the "chrom" field). For each chromosome, all its intervals are regarded as a data interval set to be queried (Table 2). We focus our analysis on annotating two types of genomic variants, of varying lengths: single nucleotide variants (SNVs) and insertions/deletions (indels), both detected in the 123,136 exome sequences and the 15,496 whole-genome sequences of the gnomAD [12]. The sizes of these population-level variant datasets are detailed per chromosome in Table 2.

Table 2 Per-chromosome sizes of genome-wide interval sets in our experiments.

| Chromosome | ENCODE intervals | gnomAD intervals |
|---|---:|---:|
| chr1 | 130,755,065 | 18,738,359 |
| chr2 | 109,795,844 | 20,204,527 |
| chr3 | 85,226,468 | 16,486,570 |
| chr4 | 62,323,066 | 16,097,457 |
| chr5 | 80,137,408 | 15,061,956 |
| chr6 | 88,374,652 | 14,016,946 |
| chr7 | 75,486,885 | 13,543,375 |
| chr8 | 61,585,063 | 12,963,423 |
| chr9 | 49,505,731 | 10,561,010 |
| chr10 | 63,396,858 | 11,139,488 |
| chr11 | 61,169,255 | 11,380,124 |
| chr12 | 66,165,580 | 10,975,733 |
| chr13 | 32,024,993 | 8,030,807 |
| chr14 | 40,938,091 | 7,574,648 |
| chr15 | 42,659,615 | 7,095,201 |
| chr16 | 46,250,642 | 7,847,535 |
| chr17 | 56,256,538 | 6,717,017 |
| chr18 | 28,192,149 | 6,353,949 |
| chr19 | 45,208,927 | 5,400,183 |
| chr20 | 37,954,390 | 5,063,026 |
| chr21 | 17,283,366 | 3,185,805 |
| chr22 | 22,644,544 | 3,245,659 |
| chrX | 36,790,451 | 9,373,753 |
| total | 1,340,125,581 | 241,056,551 |

## 4 EXPERIMENTS AND RESULTS

For each of the 23 chromosomes (22 autosomes and X chromosome), we first constructed the three tree structures based on the ENCODE interval sets. Then we performed the interval query with respect to each of Allen's relations using each interval from the gnomAD datasets as query interval. The running time was recorded for each step in order to evaluate the efficiency of the three tree structures.

Table 3 The building time (in seconds) of the three tree structures on ENCODE genomic intervals. Abbreviations: RTFC=range tree with fractional cascading; 2D-RT=basic 2-dimensional range tree; IT=interval tree.

| Chromosome | RTFC | 2D-RT | IT |
| --- | --- | --- | --- |
| chr1 | 749.23 | 1259.65 | **415.87** |
| chr2 | 606.40 | 1039.65 | **400.09** |
| chr3 | 456.99 | 680.08 | **303.82** |
| chr4 | 315.89 | 541.14 | **168.16** |
| chr5 | 426.93 | 710.28 | **246.75** |
| chr6 | 480.34 | 825.22 | **267.85** |
| chr7 | 401.38 | 580.27 | **221.49** |
| chr8 | 315.30 | 544.01 | **173.46** |
| chr9 | 249.58 | 417.13 | **135.34** |
| chr10 | 328.04 | 554.97 | **174.88** |
| chr11 | 314.78 | 532.39 | **173.40** |
| chr12 | 345.88 | 585.10 | **194.14** |
| chr13 | 147.71 | 249.00 | **77.88** |
| chr14 | 199.04 | 336.07 | **107.22** |
| chr15 | 211.76 | 351.60 | **113.49** |
| chr16 | 231.40 | 379.06 | **127.23** |
| chr17 | 294.13 | 484.06 | **145.69** |
| chr18 | 131.49 | 218.94 | **63.52** |
| chr19 | 229.74 | 371.54 | **120.93** |
| chr20 | 187.66 | 314.78 | **99.71** |
| chr21 | 75.40 | 126.74 | **37.68** |
| chr22 | 105.05 | 173.46 | **50.64** |
| chrX | 167.48 | 294.04 | **85.03** |
| total | 6971.58 | 11569.15 | **3904.27** |

We followed the range tree building algorithm `Build2DRangeTree` to build 2D-RT [16], and RTFC was built following a well-defined modification of the `Build2DRangeTree` algorithm that addressed the differences between the structures of RTFC and 2D-RT [23]. IT was built by iteratively inserting a node corresponding to a point into an initially empty tree. Since IT is an augmented RB-tree (Red-Black tree), IT insertion algorithm `Interval-Insert` is a well-defined modification of the `RB-Insert` algorithm outlined by Cormen *et al.* [24]. In our experiments, the building times for RTFC, 2D-RT and IT are shown in Table 3. From Table 3, IT took the shortest time and 2D-RT took the longest time to build the corresponding tree structures. According to Figure 4, IT has the least complex structure among the three trees. For RTFC, each node needs to additionally maintain one index array and two FC-index arrays. For a 2D-RT, each node additionally maintains a 1-dimensional range tree. Since a range tree has a more complex structure than an array, constructing a 2D-RT takes more time than constructing an RTFC. Consistently, in our experiments, we saw that IT < RTFC < 2D-RT in building time. On the other hand, we note that once the tree structure was built, it could be used for interval query regarding any relations for any given query intervals. Thus, for all queries on one chromosome, building the tree data structure was just one-time up-front effort and we focused our time complexity analysis on repeated query processes.

Since '<' and '>' queries can be transformed to $d$ queries as explained in Section 3.1, we only considered the queries regarding the 11 intersection relations in Allen's algebra in our experiments. The time complexity evaluations of these query types on all the 23 chromosomes are summarized in Table 4. From Table 4, RTFC and 2D-RT are more efficient than IT for most of the interval query types, and RTFC consistently consumes less time than 2D-RT for the 11 intersection queries. For $s$, $si$, $mi$ and $=$ queries, since the result intervals must have a fixed start, these queries mainly perform binary searches to find the tree nodes corresponding to the start position (subsequent searches are within these nodes only). This procedure is similar to query on IT with interval starts as keys, which is consistent with the observation that IT and range trees have comparable query efficiency.

Table 4 The query time of the 11 intersection queries (in seconds) and the corresponding result set sizes with gnomAD intervals as query intervals and ENCODE intervals as data intervals. Abbreviations: RTFC=range tree with fractional cascading; 2D-RT=basic 2-dimensional range tree; IT=interval tree.

| Query | RTFC | 2D-RT | IT | Result size |
|---|---|---|---|---|
| o | **97.19** | 258.18 | 1774.81 | 36,731,416 |
| oi | **26.83** | 30.57 | 128.20 | 36,573,171 |
| d | **26.63** | 29.00 | 132.00 | 96,633 |
| di | **551.69** | 986.35 | 1935.54 | 47,218,140,890 |
| s | **35.60** | 137.67 | 128.68 | 3,005 |
| si | 144.63 | 156.85 | **136.54** | 113,873,940 |
| f | **40.08** | 43.68 | 126.52 | 3,220 |
| fi | **319.12** | 554.89 | 1733.57 | 113,684,059 |
| m | **303.59** | 548.73 | 1729.51 | 113,785,360 |
| mi | **132.77** | 155.19 | 134.88 | 114,013,875 |
| = | 147.69 | 157.90 | **124.21** | 417 |

## 5 DISCUSSION

Allen's interval algebra can be applied to genomic intervals to refine the relations between two genomic intervals. There are 13 possible relations between two intervals according to Allen's interval algebra. Through rewriting the definition of Allen's interval relations, we transform the interval query problem to the range query problem and efficiently solve the problem using the augmented range tree data structure. Though there are many studies on the interval queries for the coarse intersection relation in the literature, this is the first study with implementation, to our knowledge, that tries to improve the query efficiency for more refined interval relations defined in Allen's algebra. Our results show that the range tree data structure can be more efficient than an interval tree for the genomic interval queries in most cases, and the technique of fractional cascading can further improve the query efficiency for range trees.

From the results in Table 4, we can also observe that different queries on the same tree structure can consume quite different times even though their result sizes are approximately equal, e.g., $o$ vs. $oi$, $si$ vs. $fi$, and $m$ vs. $mi$. Though the theoretical time complexities for range trees ($O(logn + k)$ for RTFC, $O(log^2n + k)$ for 2D-RT) are the same across different the query relations, the search orders for the two coordinates in our implementation and the widths of the query intervals make the actual query times different (i.e., they affect the constant factors hidden in the $O(\cdot)$ notations). In our implementation of the range tree, we built the range tree using the start of an interval as the key of a node in the first level tree (i.e., $x$-tree). During the query, we first searched the $x$-tree then searched the $y$-trees (for 2D-RT) or a point array (for RTFC). If constraint on the start is rewritten to correspond to a narrow range, we can eliminate many intervals whose starts are not in this range by the first search on the $x$-tree, and the following search on $y$-trees that have less intervals become more efficient. For example, for a query interval $[x', y']$, an $oi$ query ($x' < x < y'$) usually has a narrower range for the start than an $o$ query ($0 < x < x'$). This is consistent with our observation that $oi$ queries consume less times than $o$ queries in our implementation. Similarly, $si$ queries ($x = x'$) are more efficient than $fi$ queries ($0 < x < x'$) and $m$ queries ($0 < x < x'$) are less efficient than $mi$ queries ($x = y'$).

For simple queries where the start is fixed, such as *s*, *si*, *mi,* and =, as explained in Section 4, their major tasks are to search all the intervals that have a certain start essentially. In these cases, one can use a simple binary search tree such as an interval tree to perform the query as efficiently as a range tree, or even more efficiently due to interval tree's simpler data structure. For example, a range tree stores all the points in leaf nodes, while an interval tree can store points in internal nodes. Thus a range tree is one level higher than an interval tree when storing the same number of intervals.

The range tree data structure is more complex than the interval tree structure as shown in Figure *4*; constructing a more complex data structure usually takes more time. However, the tree building process is upfront, once the tree structure is constructed, it can be used for interval queries with respect to any relations and any query intervals repeatedly. Thus, as a general case in genomics, when performing a large number of interval queries, improving the query efficiency is much more important than improving the building efficiency. Nonetheless, we plan to improve the building efficiency for range trees by optimizing the insertion algorithm in our future study.

# 6   CONCLUSION

Allen's interval algebra can provide more refined relations between intervals. These more refined relations can provide more detailed information for genomic annotations and facilitate future discovery on genomics. Improving the efficiency of interval query regarding relations in Allen's algebra is essential for multiple bioinformatics applications. In this study, we developed a novel approach for efficient interval queries, which is optimal in theory and fast in practice. Our approach transforms an interval query problem to a range query problem by rewriting the definition of Allen's interval relations. We then developed and implemented a basic 2-dimensional range tree (2D-RT) and an augmented range tree to efficiently solve the range query problem. In particular, we built range tree with fractional cascading (RTFC), that has an optimal theoretical time complexity of $O(\log n + k)$. Our experimental results show that 2D-RT is more efficient than an interval tree for most of the queries and RTFC can further improve the query efficiency. Thus, the range tree with fractional cascading provides a data structure that can perform the interval queries efficiently on genomic datasets and can facilitate faster genomic data analysis and knowledge discovery.